\documentclass[
  twocolumn,
  preprintnumbers,
  superscriptaddress,
  reprint,
  longbibliography,
  floatfix,
  raggedbottom,
  prl
]{revtex4-2}

\usepackage{amsmath}
\usepackage{amssymb}
\usepackage{bm}
\usepackage{physics}
\usepackage{graphicx}
\usepackage{xcolor}
\usepackage[
  colorlinks=true,
  linkcolor=blue,
  citecolor=blue,
  urlcolor=blue
]{hyperref}

\begin{document}

\title{Fractionalization-Induced Time-Reversal Symmetry Breaking at Continuous Superconducting Transitions in Low-Symmetry Kondo Lattices}

\author{Wonjune Choi}
\affiliation{%
Theoretical Division, T-4 and CNLS, Los Alamos National Laboratory, Los Alamos, New Mexico 87545, USA
}

\author{Shi-Zeng Lin}
\email{szl@lanl.gov}
\affiliation{%
Theoretical Division, T-4 and CNLS, Los Alamos National Laboratory, Los Alamos, New Mexico 87545, USA
}
\affiliation{%
Center for Integrated Nanotechnologies (CINT), Los Alamos National Laboratory, Los Alamos, New Mexico 87545, USA
}

\begin{abstract}
In conventional Landau theory, simultaneous continuous onset of superconductivity and the breaking of additional symmetries generally requires a multidimensional order parameter or fine tuning.
We show that symmetry fractionalization provides a different route.
In a Kondo lattice with a $\mathbb Z_2$ spin liquid, the Kondo fields inherit the projective symmetry action of the spinons.
When the projective time-reversal action on these fields squares to $-1$, the projective symmetry algebra forbids a condensate that preserves time-reversal symmetry and enforces twofold degeneracy of the critical modes.
Because Kondo condensation in a $\mathbb Z_2$ spin liquid also breaks electromagnetic $U(1)$ symmetry, superconductivity and time-reversal symmetry breaking can emerge at a single continuous transition, even if crystal symmetry admits only one-dimensional irreducible representations.
We demonstrate this mechanism in an anisotropic Kondo--Kitaev model, where superconductivity and the breaking of time-reversal and mirror symmetries emerge simultaneously.
\end{abstract}

\maketitle

Within the conventional Ginzburg--Landau framework, two independent one-dimensional order parameters that are not related by any symmetry generally become critical at different temperatures.
Hence, their simultaneous continuous onset requires fine tuning of the system parameters.
By contrast, when
an order parameter transforms under a multidimensional irreducible
representation of the crystal point group, symmetry enforces the
degeneracy of its components, allowing them to become critical at the
same temperature.
In superconductors, this provides a familiar route to simultaneous breaking of electromagnetic $U(1)$ and time-reversal symmetry (TRS) \cite{SigristUeda1991}.
However, this conventional mechanism is unavailable when the relevant crystal
point group has only one-dimensional irreducible representations.
This raises a basic question: can superconductivity and the breaking of
another symmetry, such as TRS, emerge together at a single continuous
transition without a multidimensional irreducible representation of the
crystal point group?

This question is relevant to heavy-fermion superconductors, where experiments have produced conflicting indications of multicomponent and TRS-breaking superconductivity. For example, in UTe$_2$, early Kerr and thermodynamic measurements motivated a multicomponent TRS-breaking state \cite{Hayes2021}, whereas higher-quality samples show a single sharp superconducting transition and challenge the intrinsic origin of the TRS-breaking signatures \cite{Rosa2022,Ajeesh2023,Azari2023,Theuss2024}.
Likewise, in PrOs$_4$Sb$_{12}$, muon-spin relaxation and Kerr measurements reported broken TRS in the superconducting state \cite{Aoki2003,LevensonFalk2018}, while recent strain measurements constrain the superconducting order parameter to the one-dimensional $A_1$ representation \cite{Vianez2026}.

The prototypical model for heavy-fermion systems consists of itinerant conduction electrons coupled to localized magnetic moments through Kondo exchange.
In sufficiently frustrated systems, the local moments may form a quantum spin liquid (QSL)~\cite{SavaryBalents2017} rather than a conventional ordered state, leading to a fractionalized Fermi liquid (FL*) in which a QSL coexists with conduction electrons \cite{Senthil2003}.
In such a QSL, fractionalized spinons can transform projectively under physical symmetries, so states with identical unbroken symmetries can nevertheless belong to distinct symmetry-fractionalization classes \cite{PSG,symmetryFractionalization}.
When a $\mathbb Z_2$ QSL is coupled to conduction electrons, the spin-singlet Kondo hybridization and pairing fields contain both an ordinary charged electron and a fractionalized spinon \cite{Senthil2003,Choi2018,ChatterjeeQiSachdevSteinberg2016}.
Because their transformation combines the ordinary symmetry action on the electron with the projective action on the spinon, the parent QSL's symmetry fractionalization may determine both the symmetry of the Kondo-condensed phase and the degeneracy of its critical modes~\cite{Christos2023}.

Indeed, symmetry fractionalization can obstruct symmetry-preserving anyon condensation \cite{Bischoff2019}, an obstruction that can also arise for electron--spinon Kondo composites.
Earlier work on doped QSLs and Kondo lattices has shown how the internal gauge structure of a condensate and its transformation under physical symmetries can lead to spontaneous symmetry breaking ~\cite{YouKimchiVishwanath2012,RauKee2014,ChandraColemanFlint2013,DrechslerVojta2025}.
Rau and Kee showed that incompatible gauge structures can force TRS and spatial-symmetry breaking upon hybridization \cite{RauKee2014}.
In hastatic order, a mismatch between non-Kramers local moments and Kramers conduction states makes the hybridization itself a spinor with $\mathcal{T}^2=-1$, so that its condensation necessarily breaks TRS \cite{ChandraColemanFlint2013}.
More recently, studies of order fractionalization in Kitaev–Kondo models have linked the condensation of charge-$e$ Kondo fields to composite superconducting order and Majorana excitations ~\cite{TsvelikColeman2022,ColemanPanigrahiTsvelik2022,PanigrahiTsvelikColeman2024,ZhuangColeman2024}.

In this Letter, we establish a precise criterion for when symmetry fractionalization forces spontaneous symmetry breaking upon Kondo condensation and linking this obstruction to the protected degeneracy of the critical modes.
In particular, projective time reversal forbids a TRS-preserving condensate and enforces the emergent Kramers degeneracy of the critical Kondo fields.
Because Kondo transition from a $\mathbb Z_2$ FL* also breaks charge $U(1)$ symmetry, this projective symmetry algebra allows superconductivity and TRS breaking to emerge at a single continuous transition without fine tuning.
We numerically demonstrate this mechanism in the anisotropic Kondo--Kitaev model, whose crystal point group admits only one-dimensional irreducible representations.

\emph{Projective symmetry of the Kondo fields.---}
We consider spin-$\tfrac12$ conduction $c$ electrons coupled to spin-$\tfrac12$ local moments $\hat{\bm S}$ through an on-site antiferromagnetic Kondo interaction $J_K$. 
At $J_K=0$, the local moments are assumed to form a $\mathbb Z_2$ QSL below a temperature $T_Q$, whose low-energy excitations include deconfined fermionic spinons.
Hence, we represent the local moments using Abrikosov fermions (spinons)~\cite{Abrikosov1965,PSG},
\begin{align}
\hat{\bm S}_j
&=
\frac12
\sum_{\sigma\sigma'}
f_{j\sigma}^\dagger
\bm{\sigma}_{\sigma\sigma'}
f_{j\sigma'},
&
\sum_\sigma
f_{j\sigma}^\dagger f_{j\sigma}
&=1.
\label{eq:Abrikosov}
\end{align}
We denote the temperature of the Kondo-driven transition by $T_c$ and assume $T_c\ll T_Q$, so that the transition is approached from a finite-temperature deconfined QSL.

Within the physical one-spinon-per-site Hilbert space, the Kondo exchange can be written as
\begin{align}
J_K \hat{\bm{s}}_j \cdot\hat{\bm S}_j
=
-\frac{J_K}{4}
\left(
\hat V_j^\dagger\hat V_j
+
\hat\Delta_j^\dagger\hat\Delta_j
-
1
\right),
\label{eq:Kondo-Fierz}
\end{align}
where $\hat V_j= \sum_\sigma c_{j\sigma}^\dagger f_{j\sigma}$ and $\hat\Delta_j = c_{j\uparrow}f_{j\downarrow} - c_{j\downarrow}f_{j\uparrow}$ are the on-site spin-singlet hybridization and pairing operators, and $\hat{\bm{s}}_j = \tfrac12 \sum_{\sigma\sigma'} c_{j\sigma}^\dagger \bm{\sigma}_{\sigma\sigma'} c_{j\sigma'}$ is the conduction electron spin density.

The corresponding mean fields, $V_j = \langle \hat{V}_j \rangle$ and $\Delta_j = \langle \hat{\Delta}_j \rangle$, form the gauge-dependent Kondo doublet $h_j = ( V_j, - \Delta_j^* )^\intercal$.
Because each Kondo field contains both a conduction electron and a spinon, its transformation combines the ordinary symmetry action on the conduction electron with the projective action on the spinon.
(see End Matter for a review of the projective symmetry group (PSG)) Let $g$ denote a physical symmetry operation and $G_g(j)\in SU(2)$ its accompanying local gauge transformation on the spinons, chosen so that the combined operation $\widetilde g=G_gg$ leaves the parent spin-liquid mean-field ansatz invariant. Under this projective symmetry operation and a global electromagnetic phase rotation $c_{j\sigma}\to e^{i\theta}c_{j\sigma}$, the Kondo doublet transforms as 
\begin{align}
h_j
\to
e^{-i\theta}
G_g^\intercal\bigl(g(j)\bigr)
h_{g(j)}^{\star_g},
\end{align}
where $h_j^{\star_g} = h_j$ for unitary $g$, and $h_j^{\star_g} = h_j^*$ for antiunitary $g$.

For a translation-invariant Kondo-hybridized state, we collect all Kondo doublets in one unit cell into a column vector $\bm h$, with indices $a,b$ running over sublattices, orbitals, and other non-spin flavors.
Let $\mathcal U_g$ denote the ordinary action of $g$ on these indices, including any permutation or mixing among them.
The full transformation can be written as $\bm h \to e^{-i\theta} D_g\bm h^{\star_g}$ with a $2\times2$ block $(D_g)_{ab} = (\mathcal U_g)_{ab} G_g^\intercal(b)$, which combines the ordinary symmetry action on the non-spin labels with the projective gauge transformations.
It is unitary because both $\mathcal U_g$ and the gauge transformations $G_g$ are unitary.

A mean-field Hamiltonian with a nonzero Kondo field $\bm h$ preserves $g$, possibly combined with a global electromagnetic rotation, if a single phase $\theta_g$ satisfies
\begin{align}
e^{-i\theta_g}
D_g\bm h^{\star_g}
=
\bm h.
\label{eq:physical-invariance}
\end{align}
The phase $\theta_g$ must be the same for every component of $\bm h$.
Independent phase rotations of different sublattices, orbitals, or flavors do not represent the global electromagnetic symmetry.
Therefore, a nontrivial projective realization of a symmetry alone does not imply its spontaneous breaking.
At the onset of a continuous Kondo transition, the symmetry $g$ is broken only if no global phase $\theta_g$ satisfies Eq.~\eqref{eq:physical-invariance} for the emerging Kondo field.

\emph{Time reversal: fractionalization enforced symmetry breaking.---}
While two applications of time reversal leave the physical spin operator unchanged, $\mathcal T^2\hat{\bm S}_j\mathcal T^{-2}=\hat{\bm S}_j$, two successive projective time-reversal transformations of a deconfined spinon may differ from the identity by the nontrivial element of the invariant gauge group (IGG).
Let us focus on the nontrivial fractionalization class, whose induced action on the Kondo fields satisfies
\begin{align}
\widetilde{\mathcal T}^{\,2}
&=
\bigl[G_{\mathcal T}(j)\mathcal T\bigr]^2
=
G_{\mathcal T}(j)G_{\mathcal T}^*(j)
=
-I_2,
\quad
\forall j,
\nonumber\\
&\Rightarrow
\Theta^{2}
=
D_{\mathcal T}D_{\mathcal T}^*
=
-\bm I.
\label{eq:projective-TR-square}
\end{align}
Here $\Theta\equiv D_{\mathcal T}\mathcal K$ is the antiunitary time-reversal transformation of the Kondo fields, and $\mathcal K$ denotes complex conjugation.
The minus sign identifies the PSG quantum number $\eta_{\mathcal T,\mathcal T}=-1$, with $g_1 = g_2 = \mathcal T$ in Eq.~\eqref{eq:IGG}.

We now ask whether a finite Kondo condensate $\bm h\neq0$ can preserve TRS.
Applying Eq.~\eqref{eq:physical-invariance} twice gives
\begin{align}
\bm h
=
e^{-i\theta_{\mathcal T}}
D_{\mathcal T}
\left(
e^{-i\theta_{\mathcal T}}
D_{\mathcal T}
\bm h^*
\right)^*
=
D_{\mathcal T}D_{\mathcal T}^*
\bm h
=
-\bm h.
\label{eq:TR-obstruction}
\end{align}
The electromagnetic phases cancel because the second application complex-conjugates the phase introduced by the first.
Equation~\eqref{eq:TR-obstruction} has only the solution $\bm h=0$.
Therefore, no finite Kondo condensate with this projective symmetry action can preserve TRS.

For a continuous transition, electromagnetic $U(1)$ symmetry restricts the quadratic free energy to $\mathcal F^{(2)}=\bm h^\dagger\mathcal R\bm h$, where $\mathcal R=\mathcal R^\dagger$.
TRS of the uncondensed state requires $\mathcal R D_{\mathcal T}=D_{\mathcal T}\mathcal R^*$.
Thus, if $\mathcal R \bm u=r \bm u$ with $r\in\mathbb R$ and $\bm u^\dagger \bm u =1$, then $\mathcal R(D_{\mathcal T} \bm u^*)=rD_{\mathcal T} \bm u^*$.
Because $D_{\mathcal T}D_{\mathcal T}^*=-\bm I$, Kramers theorem gives $\bm u^\dagger \left(D_{\mathcal T}\bm u^*\right)=0$.
Therefore, the modes form degenerate Kramers pairs $(\bm u,D_{\mathcal T} \bm u^*)$, although the Kondo fields carry no physical spin indices.

If one pair becomes critical while all other modes remain massive, its quadratic part of the free energy is $\mathcal F_{\mathrm{crit}}^{(2)} = r\left( |\zeta_1|^2 + |\zeta_2|^2 \right)$, where $\bm h_{\text{crit}} = \zeta_1 \bm u + \zeta_2 D_{\mathcal T} \bm u^*$ is the critical Kondo field.
The sign of $r$ controls the instability of both $\zeta_1$ and $\zeta_2$.
Higher order terms select the relative phase of the two order parameters and determine whether the transition is continuous.

To sum up, the projective algebra pairs the critical Kondo modes,
generating an emergent multidimensional critical subspace even when the
crystal point group admits only one-dimensional irreducible
representations.
It also forbids every nonzero Kondo condensate from preserving TRS.
Thus, at a direct continuous transition from FL* to a Kondo-induced superconductor, superconductivity and TRS breaking begin together.
The End Matter gives a geometric interpretation of this obstruction.

\emph{Inversion: projective symmetry without symmetry breaking.---}
A negative projective square alone does not require spontaneous symmetry breaking.
As a unitary counterexample, consider a projective inversion that exchanges two sublattices $s=A,B$:
\begin{equation}
D_{\mathcal I}
=
\begin{pmatrix}
0&-I_2\\
I_2&0
\end{pmatrix},
\qquad
D_{\mathcal I}^2=-\bm I.
\label{eq:I-action}
\end{equation}
This corresponds to the inversion fractionalization class with $\eta_{\mathcal I,\mathcal I} = -1$.

For a finite condensate $\bm h = (h_A, h_B)^\intercal = (h, -ih)^\intercal$, $\bm h$ is an eigenvector of $D_{\mathcal I}$, i.e., $D_{\mathcal I}\bm h=i\bm h$.
Then, Eq.~\eqref{eq:physical-invariance} can be satisfied with $\theta_{\mathcal I}=\pi/2$.
Unlike time reversal, a unitary symmetry does not complex-conjugate the compensating electromagnetic phase when the operation is repeated.
So, the global charge $U(1)$ phase can cancel the projective sign in Eq.~\eqref{eq:I-action}, allowing the mean-field Hamiltonian with $\bm h\neq0$ to preserve inversion symmetry.
More generally, a unitary symmetry $g$ is preserved when $\bm h$ is an eigenvector of $D_g$, $D_g\bm h = e^{i\theta_g}\bm h$.
The eigenvalue is unimodular because $D_g$ is unitary.
If no single phase $\theta_g \in [0, 2\pi)$ satisfies Eq.~\eqref{eq:physical-invariance}, the unitary symmetry $g$ is also spontaneously broken.

\emph{Electromagnetic $U(1)$ symmetry: superconductivity induced by spinon pairing.---}
The superconducting order can be diagnosed from a gauge-invariant charge-two bond order.
Let $u_{jk}^{\mu}$ be the mean-field matrix of a spinon Hamiltonian on link $jk$; $\mu=0$ labels the singlet channel, and $\mu=x,y,z$ label the triplet channels [Eq.~\eqref{eq:Hmf}].
Then
\begin{align}
\mathcal P_{jk}^{\mu}=h_j^\intercal u_{jk}^{\mu} (i\sigma^y) h_k
\label{eq:charge-two-composite}
\end{align}
is invariant under the $SU(2)$ gauge transformation for spinons.
Under electromagnetic $U(1)$ rotation, it transforms as $\mathcal P_{jk}^{\mu}\to e^{-2i\theta}\mathcal P_{jk}^{\mu}$.
Hence, any nonzero $\mathcal P_{jk}^{\mu}$ spontaneously breaks electromagnetic $U(1)$ symmetry down to $\mathbb Z_2$ and constitutes a superconducting order parameter.

In the gauge $\bm h_j=(V_j,0)^\intercal$, the charge-two bond order reduces to $\mathcal P_{jk}^{\mu}=-V_j(u_{jk}^{\mu})_{12}V_k$.
This gauge-fixed expression makes explicit how an anomalous spinon-link amplitude combines with the Kondo condensates to produce Cooper pairing of the conduction electrons~\cite{ColemanAndrei1989,Senthil2003}.
When both components of the Kondo doublet are nonzero, both normal and anomalous matrix elements of $u_{jk}^{\mu}$ can contribute to $\mathcal P_{jk}^{\mu}$.

\emph{Kondo--Kitaev model.---}
As a concrete example of symmetry breaking induced by fractionalization, we consider the Kondo--Kitaev model on a bilayer honeycomb lattice~\cite{Seifert2018,Choi2018,LundemoSudbo2024,Bunney2025}:
\begin{multline}
\hat{H}
=
-\sum_{\langle j,k\rangle_\alpha}\sum_{\sigma}
t_\alpha
\left(
c_{j\sigma}^\dagger c_{k\sigma}
+
c_{k\sigma}^\dagger c_{j\sigma}
\right)
\\
-\sum_{\langle j,k\rangle_\alpha}
K_\alpha \hat{S}_j^\alpha \hat{S}_k^\alpha
+
J_K\sum_j
\hat{\bm s}_j\cdot\hat{\bm S}_j ,
\end{multline}
where $\langle j,k\rangle_\alpha$ denotes an $\alpha$-type link of the honeycomb lattice \cite{Kitaev2006}.
For $t_x=t_y>t_z$ and $K_x=K_y>K_z$, the threefold rotational symmetry is explicitly broken, and the remaining crystal symmetry is described by the point group $C_{2h}$, which admits only one-dimensional representations.
Hence, the conventional Ginzburg--Landau framework requires fine tuning for two independent order parameters to have simultaneous continuous onset.

In the thermodynamic limit, a strictly two-dimensional QSL is stable only at zero temperature \cite{CastelnovoChamon2007,Nasu2014}, precluding the finite-temperature phase transition considered here.
By contrast, three-dimensional Kitaev models on hyperhoneycomb or harmonic-honeycomb lattices~\cite{MandalSurendran2009} can support a deconfined phase at finite temperatures, with $T_Q>0$~\cite{Nasu2014,Kimchi2014}.
Hence, we employ the two-dimensional Kondo--Kitaev model as a numerically tractable demonstration at the mean-field level.
A physical realization of this Kondo scenario requires a three-dimensional $\mathbb Z_2$ QSL that is Kondo-coupled to conduction electrons.


\begin{figure*}[t]
    \centering
    \includegraphics[width=\textwidth]{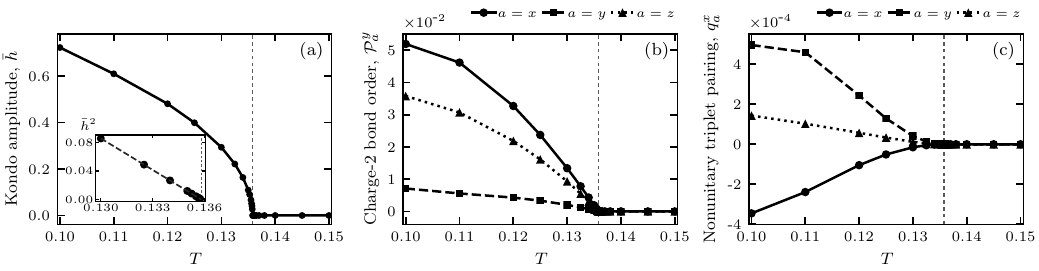}
    \setlength{\abovecaptionskip}{1pt}
    \vspace{-8pt}
    \caption{
    The panels show mean-field results for the Kondo--Kitaev model on $20\times20$ unit cells with periodic boundary conditions, $K_x = K_y = 1$, $K_z=0.95$, $t_x = t_y = 0.2$, $t_z=0.19$, $J_K=1.9$, and conduction electron filling $n_c=0.9$ per site.
(a) The mean-field scaling $\bar h^2 \propto T_c - T$ below $T_c$ (inset) supports continuous onset of the Kondo amplitude $\bar h$.
(b) Charge-two bond order $\mathcal P_a^y$ indicates Kondo-induced superconductivity.
(c) Nonunitary triplet pairing $q_a^x$ extracted from the anomalous $c$ fermion correlators diagnoses broken TRS. Within numerical resolution, all three order parameters become nonzero continuously at the same critical temperature $T_c \simeq 0.1358$.
}
    \label{fig:order_params}
\end{figure*}

The static $\mathbb{Z}_2$ gauge-field background of the Kitaev model determines the PSG of its anisotropic phase~\cite{YouKimchiVishwanath2012,Choi2018}.
Time reversal $\mathcal T$, inversion $\mathcal I$ about a $z$-bond center, and mirror reflection $\sigma$ across the line through the $z$-bond each require a nontrivial gauge transformation to leave the ansatz invariant.
All three have fractionalization quantum number $-1$: $\eta_{\mathcal T, \mathcal T} = \eta_{\mathcal I, \mathcal I} = \eta_{\sigma,\sigma} = -1$.
As shown above, this time-reversal fractionalization forces TRS breaking at a continuous Kondo-hybridization onset.
Inversion and mirror symmetries are preserved only when their projective actions satisfy Eq.~\eqref{eq:physical-invariance}.

We solve the model at finite temperature using Abrikosov-fermion mean-field theory~\cite{BurnellNayak2011,SupplementalMaterial}.
The approximation includes a thermally averaged, self-consistent gauge-field background but neglects vison proliferation.
It is appropriate for three-dimensional topological order at temperatures well below the vison gap.

For the parameters in Fig.~\ref{fig:order_params}, the self-consistent solutions are well described by
\begin{align}
h_A(T) = i h_B(T) =\bar{h}(T) \frac{1+i}{2}
\begin{pmatrix}
1 \\ -1
\end{pmatrix},
\label{eq:scf_sol}
\end{align}
where the Kondo amplitude $\bar h = \|h_A \| = \|h_B\|$.
The smooth onset of $\bar h$ and the near-linear dependence of $\bar h^2$ on temperature near $T_c \simeq 0.1358$ support a continuous mean-field transition [Fig.~\ref{fig:order_params}(a)].
The charge-two bond order $\mathcal P_a^y$ confirms superconductivity from Kondo hybridization and intrinsic pairing of charge-neutral spinons [Fig.~\ref{fig:order_params}(b)].

We also compute the anomalous $c$-fermion correlators on each bond type,
\begin{align}
\Phi_a^{\sigma\sigma'}
=
\ev{c_{j\sigma}c_{k\sigma'}}
\equiv
\left(
\psi_a+\vb d_a\cdot\bm\sigma
\right)i\sigma_y,
\end{align}
where $\ev{j,k}_a$ denotes an $a$-type bond.
The superconducting state is nonunitary and breaks TRS when the nonunitary triplet pairing vector on each bond, $\vb q_a = i\,\vb d_a\times\vb d_a^*$, is nonzero~\cite{SigristUeda1991}.
Figure~\ref{fig:order_params}~(c) shows that $\vb q_a=(q_a^x,0,0)\neq\vb 0$ for $T<T_c$.
This result supports simultaneous breaking of electromagnetic $U(1)$ symmetry and TRS at $T_c \simeq 0.1358$ in a low-symmetry crystal, where Ginzburg--Landau theory alone requires fine tuning for the two onsets to coincide.

Equation~\eqref{eq:I-action} gives the projective inversion action in the Kitaev honeycomb model.
The translation-invariant field $\bm h=(h_A,h_B)^\intercal$ can preserve inversion symmetry through the compensating electromagnetic phase in Eq.~\eqref{eq:physical-invariance}.
The lowest-free-energy solutions satisfy $h_A=i h_B$, consistent with this condition, and have equal conduction-electron densities on both sublattices, $n_A=n_B$.
Thus, the superconducting state preserves inversion symmetry.

Unlike inversion, the mean-field solution spontaneously breaks mirror symmetry.
Mirror reflection acts projectively on each Kondo doublet as
\begin{align}
h_s &\to W_\sigma h_s,
&
W_\sigma
&=
\frac{i}{\sqrt{2}}
\begin{pmatrix}
0 & 1-i\\
1+i & 0
\end{pmatrix}.
\end{align}
Substitution of the self-consistent solution in Eq.~\eqref{eq:scf_sol} shows that no single electromagnetic phase $\theta_\sigma$ satisfies $W_\sigma h_s=e^{i\theta_\sigma}h_s$ on both sublattices.
The bond-dependent order parameters in Fig.~\ref{fig:order_params} (b) and (c) provide an independent physical diagnosis.
Mirror invariance would require $\mathcal P_x^y=\mathcal P_y^x$ and $q_x^x=-q_y^y$.
Instead, $\mathcal P_a^x=q_a^y=0$ on all three bonds, while $\mathcal P_x^y$ and $q_x^x$ are finite in Fig. \ref{fig:order_params}.
Hence, the Kondo-hybridized phase breaks mirror symmetry.

\emph{Conclusion.---}
We have shown that the degeneracy enabling simultaneous superconductivity and TRS breaking can originate from the symmetry fractionalization of a parent $\mathbb Z_2$ QSL.
The projective time-reversal algebra protects the degeneracy of the critical Kondo modes and forbids a condensate that preserves TRS, locking the two onsets at a direct continuous transition without fine tuning.
Our mean-field calculations in the anisotropic Kondo--Kitaev model support a single continuous transition into a superconducting state that breaks time-reversal and mirror symmetries while preserving inversion symmetry.

The mechanism extends to unitary symmetries with nontrivial mutual symmetry fractionalization \cite{Balents2005,Schuler2023,SongZhang2023}.
For example, physically commuting symmetries may satisfy $D_{g_1}D_{g_2}=-D_{g_2}D_{g_1}$.
This projective symmetry algebra has no one-dimensional representation and protects a multidimensional critical Kondo multiplet.
No nonzero condensate can be a simultaneous eigenstate of both symmetry operations, so at least one physical symmetry must be spontaneously broken.
The multidimensional structure enabling simultaneous symmetry breaking can originate from symmetry fractionalization, even when ordinary crystal symmetry admits only one-dimensional irreducible representations. More broadly, our work offers a new perspective for reconciling seemingly conflicting experimental observations in unconventional superconductors, where signatures of additional symmetry breaking coexist with evidence for a single superconducting transition in low-symmetry crystals.

\emph{AI usage disclosure.---}
OpenAI Codex (GPT-5.6) assisted with writing numerical code for the mean-field calculations, plotting data, and polishing the manuscript.
The authors directed these uses, reviewed the code, independently analyzed the data, and checked the resulting plots and text.
The authors take responsibility for the content of the manuscript.

\begin{acknowledgments}
{\it Acknowledgments.---} The authors thank Yahui Zhang and Hao Zhang for helpful discussions.
The work at LANL was carried out under the auspices of the U.S. DOE NNSA under contract No. 89233218CNA000001 through the LDRD Program, and was performed, in part, at the Center for Integrated Nanotechnologies, an Office of Science User Facility operated for the U.S. DOE Office of Science, under user proposals $\#2018BU0010$ and $\#2018BU0083$.

\end{acknowledgments}

\bibliography{Kondo}

\clearpage
\onecolumngrid
\section*{End Matter}

\twocolumngrid
\appendix
\emph{Projective symmetry group.---}
The Abrikosov-fermion (spinon) representation in Eq.~\eqref{eq:Abrikosov} is redundant because different fermionic descriptions can represent the same physical spin state.
To display this redundancy, we arrange the spinons into the matrix
\begin{align}
F_j
&=
\begin{pmatrix}
f_{j\uparrow} & f_{j\downarrow}^\dagger \\
f_{j\downarrow} & -f_{j\uparrow}^\dagger
\end{pmatrix},
&
\hat{\bm S}_j
&=
\frac14
\tr\left(
F_j^\dagger
\bm{\sigma}
F_j
\right).
\label{eq:Fmatrix}
\end{align}
Since $F_j\to F_jW_j$ leaves $\hat{\bm S}_j$ unchanged for any $W_j\in SU(2)$, the representation has a local $SU(2)$ gauge redundancy~\cite{AffleckZouHsuAnderson1988}.

This redundancy permits each physical symmetry to be accompanied by a local gauge transformation.
For a unitary symmetry $g$, the combined transformation acts as
\begin{align}
F_j
\to
R_g^\dagger
F_{g(j)}
G_g\bigl(g(j)\bigr),
\qquad
G_g(j)\in SU(2),
\end{align}
where $R_g$ is the physical spin rotation associated with $g$ and $G_g(j)$ is the accompanying local gauge transformation.
For an antiunitary symmetry such as time reversal, the transformation also involves complex conjugation.

At the mean-field level, the interacting spinons are described by a self-consistent quadratic Hamiltonian $\hat H_{\rm MF}$.
For the Kondo lattice model, the mean-field Hamiltonian is parametrized by the $2\times2$ mean-field matrices $v_j$ and $u_{jk}^\mu$ such that
\begin{multline}
\hat{H}_{\text{MF}} = \sum_{jk,\sigma} t_{jk} c_{j\sigma}^\dagger c_{k\sigma}
+ \sum_j \tr\left(C_j v_j F_j^\dagger\right)
\\
+ \sum_{jk,\mu} \tr\left(\sigma^\mu F_j u_{jk}^\mu F_k^\dagger\right),
\label{eq:Hmf}
\end{multline}
where $C_j$ denotes the $2\times2$ matrix of conduction-electron creation and annihilation operators arranged as in Eq.~\eqref{eq:Fmatrix}.

A mean-field state preserves a symmetry $g$ if one can choose a gauge transformation $G_g$ such that the combined operation $\widetilde g=G_gg$ leaves $\hat H_{\rm MF}$ invariant.
All such combined operations constitute the projective symmetry group (PSG)~\cite{PSG}.
The choice of $G_g$ is not unique because $\widetilde g$ may be multiplied by a pure gauge transformation that already leaves $\hat H_{\rm MF}$ invariant.
These pure gauge transformations form the invariant gauge group (IGG), which is $\mathrm{IGG}=\{\pm I_2\}\cong\mathbb Z_2$ for a $\mathbb Z_2$ QSL.
This ambiguity appears in the composition of two combined symmetry operations:
\begin{align}
\widetilde g_1\widetilde g_2
=
\eta_{g_1,g_2}\widetilde{g_1g_2},
\qquad
\eta_{g_1,g_2}\in\mathrm{IGG}.
\label{eq:IGG}
\end{align}

The factors $\eta_{g_1,g_2}$ form a factor set, and gauge-equivalent factor sets describe the same projective symmetry action on the spinons.
Their gauge-equivalence classes specify the symmetry-fractionalization classes.
For the $\mathbb Z_2$ IGG, gauge-invariant signs associated with specified physical group relations are the corresponding symmetry-fractionalization quantum numbers~\cite{symmetryFractionalization}.
These quantum numbers remain well defined beyond mean-field theory as long as the spinons remain deconfined.

\emph{Geometry of projective time-reversal symmetry.---}
We illustrate how two distinct TRS fractionalization classes constrain the Kondo fields.

First, take $G_{\mathcal T}=i\sigma^y$ as an example of the nontrivial projective time-reversal action, with $G_{\mathcal T}G_{\mathcal T}^*=-I_2$.
The Kondo doublet transforms as $h_j=(V_j,-\Delta_j^*)^\intercal\to(\Delta_j,V_j^*)^\intercal$.
No nonzero doublet is invariant under this transformation, even after a global $U(1)$ charge rotation.
Thus, any nonzero Kondo condensate breaks TRS.
This includes finite hybridization $V_j\neq0$ with $\Delta_j=0$.

The obstruction to preserving TRS has a geometric interpretation in terms of the pseudospin $\mathcal S_j=h_j^\dagger\boldsymbol{\sigma}h_j$, which transforms $\mathcal S_j\to-\mathcal S_j$ under $\widetilde{\mathcal{T}} = G_{\mathcal T} \mathcal T$.
The identity $|\mathcal S_j|^2=(h_j^\dagger h_j)^2$ ensures that every nonzero Kondo field has a nonzero pseudospin vector.
Since the pseudospin is invariant under electromagnetic phase rotations, no charge rotation can compensate its sign reversal.

By contrast, take $G_{\mathcal T}=i\sigma^x$ as an example of the trivial projective time-reversal action, with $G_{\mathcal T}G_{\mathcal T}^*=+I_2$.
Its action is $h_j\to i(-\Delta_j,V_j^*)^\intercal$.
The uniform condensate $h_j=(\zeta,\zeta)^\intercal$ spans an invariant one-dimensional complex subspace, with amplitude transformation $\zeta\to i\zeta^*$.
Choosing the global charge phase to make $\zeta$ real gives $h_j\to i h_j$ under time reversal.
Then, a charge rotation with $\theta_{\mathcal T}=\pi/2$ restores the condensate after the time-reversal.
Thus, with a $\mathbb Z_2$ QSL, symmetry permits a superconducting instability described by a single complex amplitude $\zeta$ without requiring TRS breaking.

The pseudospin transformation makes this distinction explicit: $(\mathcal S_j^x,\mathcal S_j^y,\mathcal S_j^z)\to(\mathcal S_j^x,\mathcal S_j^y,-\mathcal S_j^z)$ under the trivial projective time-reversal action.
Because the pseudospin vectors in the $xy$ plane are invariant, the uniform condensate $h_j=(\zeta,\zeta)^\intercal$, which gives $\mathcal S_j=(2|\zeta|^2,0,0)$, can remain finite without breaking TRS.

\end{document}


\title{Supplemental Material for ``Fractionalization-Induced Time-Reversal Symmetry Breaking at Continuous Superconducting Transitions in Low-Symmetry Kondo Lattices''}
\author{Wonjune Choi}
\affiliation{Theoretical Division, T-4 and CNLS, Los Alamos National Laboratory, Los Alamos, New Mexico 87545, USA}
\author{Shi-Zeng Lin}
\email{szl@lanl.gov}
\affiliation{Theoretical Division, T-4 and CNLS, Los Alamos National Laboratory, Los Alamos, New Mexico 87545, USA}
\affiliation{Center for Integrated Nanotechnologies (CINT), Los Alamos National Laboratory, Los Alamos, New Mexico 87545, USA}
\maketitle
\renewcommand{\theequation}{S\arabic{equation}}

\section{Model}
\label{sec:model}

We consider an \(AA\)-stacked bilayer honeycomb lattice.
The top layer hosts noninteracting spin-\(\tfrac{1}{2}\) conduction electrons created by \(c_{j\sigma}^\dagger\), where \(\sigma=\uparrow,\downarrow\).
Throughout the Supplemental Material, repeated physical spin labels such as \(\sigma,\sigma'\) are summed in compact operator expressions unless stated otherwise; all non-spin sums are written explicitly.

The grand-canonical Hamiltonian for itinerant electrons in the top layer is
\begin{align}
\hat{K}_c
\equiv
\hat H_t -\mu \hat{N}_c
&=
-\sum_{a=x,y,z}
\sum_{\langle j,k\rangle_a}
t_a
\left(
c_{j\sigma}^\dagger c_{k\sigma}
+
c_{k\sigma}^\dagger c_{j\sigma}
\right)
-\mu\sum_j c_{j\sigma}^\dagger c_{j\sigma},
\qquad
t_a > 0 .
\end{align}
Here \(\langle j,k\rangle_a\) denotes a nearest-neighbor bond of type \(a=x,y,z\) on the honeycomb lattice.
The bottom layer hosts spin-\(\tfrac{1}{2}\) local moments interacting through bond-dependent Ising couplings, namely the Kitaev interaction,
\begin{align}
\hat H_K
=
-\sum_{a=x,y,z}\sum_{\langle j,k\rangle_a}
K_a \hat S_j^a \hat S_k^a,
\qquad
K_a > 0 .
\end{align}
The two layers are coupled through an onsite antiferromagnetic Kondo interaction,
\begin{align}
\hat H_{\mathrm{Kondo}}
=
J_K\sum_j \hat{\bm s}_j\cdot \hat{\bm S}_j,
\qquad
J_K>0,
\end{align}
where the conduction-electron spin operator is \(\hat{\bm s}_j = \frac12 c_{j\sigma}^\dagger \bm\sigma_{\sigma\sigma'} c_{j\sigma'}.\)
Here \(\bm\sigma=(\sigma^x,\sigma^y,\sigma^z)\) denotes the Pauli matrices.
The full microscopic grand-canonical Hamiltonian is \(\hat K=\hat K_c+\hat H_K+\hat H_{\mathrm{Kondo}}.\) We set $k_B=1$ and measure energies and temperatures in units of
$K_x=K_y$.

\section{Fractionalized Fermi liquid (FL*)}
\label{sec:FLstar}

A useful reference state is the mean-field solution in which Kondo hybridization is absent, so the conduction electrons and local moments are decoupled at the mean-field level.
This state is a fractionalized Fermi liquid (FL*): the Kitaev spin liquid coexists with a conduction-electron Fermi liquid.

We are interested in finite-temperature phase transitions between the FL* state and a Kondo-hybridized superconducting state.
The goal is to examine whether Kondo hybridization can yield a time-reversal symmetry broken superconducting state via a single continuous transition without fine tuning.

\subsection{Quantum order of the Kitaev spin liquid}

The quantum order of the Kitaev spin liquid constrains its fractionalized excitations.
In particular, the fermionic fractionalized excitation is sharply defined and does not mix with the vortex excitations, or visons, of the emergent \(\mathbb{Z}_2\) gauge field.
This separation can be seen explicitly as follows.

In terms of Majorana fermions \(\chi_j^\mu\) with \(\mu=0,x,y,z\), we represent the spin operator in the enlarged Hilbert space as
\begin{align}
\hat S_j^\alpha
=
\frac{i}{2}\chi_j^\alpha \chi_j^0,
\qquad
\alpha=x,y,z,
\end{align}
where
\begin{align}
\{\chi_j^\mu,\chi_k^\nu\}
=
2\delta^{\mu\nu}\delta_{jk},
\qquad
\big(\chi_j^{\mu}\big)^\dagger=\chi_j^\mu .
\end{align}
The four Majorana fermions span a four-dimensional local Hilbert space, whereas a physical spin-\(\tfrac12\) has only two states.
The physical subspace is selected by the local gauge constraint
\begin{align}
D_j
\equiv
\chi_j^x\chi_j^y\chi_j^z\chi_j^0,
\qquad
D_j\ket{\mathrm{phys}}=\ket{\mathrm{phys}}.
\label{eq:majorana-physical-constraint}
\end{align}
This constraint is necessary for the spin algebra.
For example,
\begin{align}
\left[\hat S_j^x,\hat S_j^y\right]
&=
\frac12\chi_j^x\chi_j^y
=
-\frac12\chi_j^z\chi_j^0
=
i\hat S_j^z,
\end{align}
where the second equality follows from \(D_j=1\).
Cyclic permutations give the other commutators.

Then, the bond-directional interaction of the Kitaev model can be written as
\begin{align}
\sum_{\langle j,k\rangle_\alpha}
\hat S_j^\alpha \hat S_k^\alpha
=
-\frac{1}{4}
\sum_{\langle j,k\rangle_\alpha}
\left(i\chi_j^\alpha \chi_k^\alpha\right)
\left(i\chi_j^0 \chi_k^0\right).
\end{align}

In the exactly solvable Kitaev model, the Hamiltonian separates into
itinerant matter Majoranas moving in a static $\mathbb Z_2$ gauge
background, with distinct flux sectors. We assume that the
Kondo-driven transition occurs well below the deconfinement scale, $T_c \ll T_Q$, so that in the vicinity of $T_c$ the system remains in the
deconfined spin-liquid regime. We therefore restrict the mean-field
calculation to saddles continuously connected to the flux-free
Kitaev state and neglect thermal vison proliferation.
For \(T<T_Q\), we use the mean-field Ansatz
\begin{align}
\left(i\chi_j^\alpha \chi_k^\alpha\right)
\left(i\chi_j^0 \chi_k^0\right)
\approx&
\left\langle i\chi_j^\alpha \chi_k^\alpha \right\rangle
i\chi_j^0 \chi_k^0
+
\left\langle i\chi_j^0 \chi_k^0 \right\rangle
i\chi_j^\alpha \chi_k^\alpha
-
\left\langle i\chi_j^0 \chi_k^0 \right\rangle
\left\langle i\chi_j^\alpha \chi_k^\alpha \right\rangle .
\label{eq:kitaev-majorana-bond-derivation}
\end{align}
A generic Hartree-Fock-Bogoliubov decoupling would also allow cross terms such as
\[
\langle \chi_j^\alpha \chi_k^0\rangle \chi_j^0 \chi_k^\alpha.
\]
Here we exclude these terms by choosing a mean-field Ansatz that preserves the separation between the matter Majorana sector and the emergent \(\mathbb{Z}_2\) gauge sector.

This finite-temperature topological-order argument does not apply to a strictly two-dimensional system, because \(\mathbb{Z}_2\) topological order is not stable at finite temperature; in that case \(T_Q=0\).
The mean-field study below can still give thermodynamic insight for \(d>2\), where the quantum order can remain stable at finite temperatures \(T<T_Q\).
Although the calculation below uses a two-dimensional lattice, it should be viewed as a demonstration of the mechanism.
A thermodynamically stable finite-temperature transition of this type should ultimately be discussed in three-dimensional systems.

\subsection{Abrikosov fermion representation of the Kitaev model}

We represent the local moments by Abrikosov fermions,
\begin{align}
\hat{\bm S}_j
=
\frac{1}{2}
f_{j\sigma}^{\dagger}
\bm\sigma_{\sigma\sigma'}
f_{j\sigma'},
\end{align}
subject to the one-spinon-per-site constraint
\begin{align}
\sum_{\sigma=\uparrow,\downarrow}
f_{j\sigma}^{\dagger} f_{j\sigma}=1
\qquad
\text{for every site }j.
\label{eq:one-spinon-constraint}
\end{align}
This fermionic representation enlarges the Hilbert space and introduces an internal local gauge redundancy.
At the mean-field level, it also introduces a Fierz ambiguity: quartic interactions can be decoupled into different channels.
Once a restricted set of variational channels is chosen, the relative weights assigned to those channels are not fixed by a unique universal principle.
For this reason, each mean-field Hamiltonian below should be read together with its variational channels and normalization convention.

To reproduce the correct excitation bandwidth and the exact ground state energy in the fixed-flux sector, we start from the Majorana representation of the spin operator, which permits the exact solution of the Kitaev model.
Then, we rewrite the resulting spin-liquid Hamiltonian in the Abrikosov-fermion basis.
This procedure fixes the normalization of the spin-liquid mean-field Hamiltonian and avoids the Fierz ambiguity in the spin-liquid sector.

Let us define
\begin{align}
f_{j\uparrow}
=
\frac{1}{2}\left(\chi_j^z+i\chi_j^0\right),
\qquad
f_{j\downarrow}
=
\frac{1}{2}\left(\chi_j^x+i\chi_j^y\right).
\end{align}
Equivalently,
\begin{align}
\chi_j^z
=
f_{j\uparrow}+f_{j\uparrow}^\dagger,
\qquad
\chi_j^0
=
i\left(f_{j\uparrow}^\dagger-f_{j\uparrow}\right),
\end{align}
and
\begin{align}
\chi_j^x
=
f_{j\downarrow}+f_{j\downarrow}^\dagger,
\qquad
\chi_j^y
=
i\left(f_{j\downarrow}^\dagger-f_{j\downarrow}\right).
\end{align}
These definitions make the equivalence of the two physical-space constraints explicit.
If \(n_{j\sigma}=f_{j\sigma}^\dagger f_{j\sigma}\), then
\begin{align}
i\chi_j^z\chi_j^0
=2n_{j\uparrow}-1,
\quad
i\chi_j^x\chi_j^y
=2n_{j\downarrow}-1,
\end{align}
and therefore
\begin{align}
D_j
&=
-\left(2n_{j\downarrow}-1\right)
\left(2n_{j\uparrow}-1\right).
\end{align}
The condition \(D_j=1\) selects precisely the two singly occupied states, and is thus equivalent to Eq.~\eqref{eq:one-spinon-constraint}.

We choose the bond orientation from sublattice \(0\) in unit cell \(\vb R\) to sublattice \(1\) in unit cell \(\vb R+\bm{\delta}_a\).
The bond vectors are chosen as
\begin{align}
\bm{\delta}_x=\vb a_1,
\qquad
\bm{\delta}_y=\vb a_2,
\qquad
\bm{\delta}_z=\vb 0,
\end{align}
where \(\vb a_1\) and \(\vb a_2\) are the primitive lattice vectors of the honeycomb Bravais lattice.
The site index is written as \(j=(\vb R,s)\), where \(\vb R\) is a unit-cell coordinate and \(s=0,1\) is the sublattice index.

On this oriented bond, we define
\begin{align}
A_a
&\equiv
\left\langle
i\chi^0_{\vb R,0}
\chi^0_{\vb R+\bm{\delta}_a,1}
\right\rangle,
\\
B_a
&\equiv
\left\langle
i\chi^a_{\vb R,0}
\chi^a_{\vb R+\bm{\delta}_a,1}
\right\rangle .
\end{align}
Each oriented-bond operator \(i\chi_j^\mu\chi_k^\mu\) is Hermitian, since
\begin{align}
\left(i\chi_j^\mu\chi_k^\mu\right)^\dagger
=
-i\chi_k^\mu\chi_j^\mu
=
i\chi_j^\mu\chi_k^\mu
\qquad (j\ne k).
\end{align}
Consequently, all \(A_a\) and \(B_a\) are real.
Applying the mean-field decoupling above to the bond identity in Eq.~\eqref{eq:kitaev-majorana-bond-derivation} on every oriented bond gives the real-space mean-field Hamiltonian
\begin{align}
\hat H_f
&=
\sum_{\vb R}
\sum_{a=x,y,z}
\frac{K_a}{4}
\Bigl[
B_a\,i\chi^0_{\vb R,0}
\chi^0_{\vb R+\bm\delta_a,1}
+A_a\,i\chi^a_{\vb R,0}
\chi^a_{\vb R+\bm\delta_a,1}
-A_aB_a
\Bigr].
\label{eq:Hf-real-space-majorana}
\end{align}

For later use, let \(j=(\vb R,0)\) and \(k=(\vb R+\bm\delta_a,1)\).
Direct substitution of the complex-fermion definitions gives
\begin{align}
i\chi_j^0\chi_k^0
&=
-i f_{j\uparrow}f_{k\uparrow}
+i f_{j\uparrow}^\dagger f_{k\uparrow}
+\mathrm{h.c.},
\label{eq:chi0-bond-abrikosov}
\\
i\chi_j^z\chi_k^z
&=
i f_{j\uparrow}f_{k\uparrow}
+i f_{j\uparrow}^\dagger f_{k\uparrow}
+\mathrm{h.c.},
\label{eq:chiz-bond-abrikosov}
\\
i\chi_j^x\chi_k^x
&=
i f_{j\downarrow}f_{k\downarrow}
+i f_{j\downarrow}^\dagger f_{k\downarrow}
+\mathrm{h.c.},
\label{eq:chix-bond-abrikosov}
\\
i\chi_j^y\chi_k^y
&=
-i f_{j\downarrow}f_{k\downarrow}
+i f_{j\downarrow}^\dagger f_{k\downarrow}
+\mathrm{h.c.}.
\label{eq:chiy-bond-abrikosov}
\end{align}

The corresponding spin-liquid mean-field Hamiltonian in the Abrikosov-fermion basis is
\begin{align}
\hat H_f
=
\frac{1}{2}
\sum_{\vb k,\sigma}
\psi^{f\dagger}_{\vb k,\sigma}
\mathcal H^f_{\vb k,\sigma}
\psi^f_{\vb k,\sigma}
+N_0
\varepsilon_f,
\label{eq:Hf}
\end{align}
where \(N_0=L_1L_2\) is the number of unit cells and
\begin{align}
\varepsilon_f = -\sum_{a=x,y,z} \frac{K_a}4 A_a B_a .
\end{align}
The Nambu spinor is defined as
\begin{align}
\psi^f_{\vb k,\sigma}
=
\left(
f_{\vb k,0,\sigma},
f_{\vb k,1,\sigma},
f_{-\vb k,0,\sigma}^\dagger,
f_{-\vb k,1,\sigma}^\dagger
\right)^{\intercal}
\label{eq:psi}
\end{align}
with the discrete Fourier transformation
\begin{align}
f_{\vb R,s,\sigma}
=
\frac{1}{\sqrt{N_0}}
\sum_{\vb k}
e^{-i\vb k\cdot\vb R}
f_{\vb k,s,\sigma} .
\end{align}
For the oriented bond, the hopping and pairing bilinears are
\begin{align}
\sum_{\vb R}
f_{\vb R,0,\sigma}^\dagger
f_{\vb R+\bm\delta_a,1,\sigma}
&=
\sum_{\vb k}
e^{-i\vb k\cdot\bm\delta_a}
f_{\vb k,0,\sigma}^\dagger f_{\vb k,1,\sigma},
\label{eq:fourier-oriented-hopping}
\\
\sum_{\vb R}
f_{\vb R,0,\sigma}
f_{\vb R+\bm\delta_a,1,\sigma}
&=
\sum_{\vb k}
e^{-i\vb k\cdot\bm\delta_a}
f_{-\vb k,0,\sigma}f_{\vb k,1,\sigma}.
\label{eq:fourier-oriented-pairing}
\end{align}

The spin-up block of the Hamiltonian is
\begin{align}
\mathcal H^f_{\vb k,\uparrow}
=
\begin{pmatrix}
0
&
\Delta^z+\Delta^0_{\vb k}
&
0
&
\Delta^z-\Delta^0_{\vb k}
\\
\left(\Delta^z+\Delta^0_{\vb k}\right)^*
&
0
&
\left(\Delta^z-\Delta^0_{\vb k}\right)^*
&
0
\\
0
&
\Delta^z-\Delta^0_{\vb k}
&
0
&
\Delta^z+\Delta^0_{\vb k}
\\
\left(\Delta^z-\Delta^0_{\vb k}\right)^*
&
0
&
\left(\Delta^z+\Delta^0_{\vb k}\right)^*
&
0
\end{pmatrix},
\label{eq:Hf-up}
\end{align}
Here \(\Delta^z\equiv \Delta^z_{\vb k}\) is independent of \(\vb k\) because \(\bm{\delta}_z=\vb 0\).
Similarly, the spin-down block is
\begin{align}
\mathcal H^f_{\vb k,\downarrow}
=
\begin{pmatrix}
0
&
\Delta^x_{\vb k}+\Delta^y_{\vb k}
&
0
&
\Delta^x_{\vb k}-\Delta^y_{\vb k}
\\
\left(\Delta^x_{\vb k}+\Delta^y_{\vb k}\right)^*
&
0
&
\left(\Delta^x_{\vb k}-\Delta^y_{\vb k}\right)^*
&
0
\\
0
&
\Delta^x_{\vb k}-\Delta^y_{\vb k}
&
0
&
\Delta^x_{\vb k}+\Delta^y_{\vb k}
\\
\left(\Delta^x_{\vb k}-\Delta^y_{\vb k}\right)^*
&
0
&
\left(\Delta^x_{\vb k}+\Delta^y_{\vb k}\right)^*
&
0
\end{pmatrix}.
\label{eq:Hf-down}
\end{align}
Here
\begin{align}
\Delta_{\vb k}^a
&=
\frac{iK_a}{4}
A_a
e^{-i\vb k\cdot\bm{\delta}_a},
\qquad
a=x,y,z,
\label{eq:scf_Kitaev_a}
\\
\Delta^0_{\vb k}
&=
\sum_{a=x,y,z}
\frac{iK_a}{4}
B_a
e^{-i\vb k\cdot\bm{\delta}_a}.
\label{eq:scf_Kitaev_0}
\end{align}
Because \(A_a\) and \(B_a\) are real,
\begin{align}
\Delta^\nu_{-\vb k}
=
-\left(\Delta^\nu_{\vb k}\right)^*,
\qquad
\nu=0,x,y,z.
\label{eq:spinon-delta-particle-hole-relation}
\end{align}
Thus the displayed matrices are Hermitian and obey the BdG particle--hole relation
\begin{align}
\tau_x
\left(\mathcal H^f_{-\vb k,\sigma}\right)^*
\tau_x
=
-\mathcal H^f_{\vb k,\sigma},
\qquad
\tau_x=
\begin{pmatrix}
0&I_2\\
I_2&0
\end{pmatrix}.
\label{eq:spinon-BdG-particle-hole}
\end{align}

With the reciprocal lattice vectors \(\vb b_1,\vb b_2\) satisfying \(\vb a_i\cdot\vb b_j=2\pi\delta_{ij},\) a physical crystal momentum is written in reduced reciprocal coordinates as
\begin{align}
\vb k=k_1\vb b_1+k_2\vb b_2,
\qquad
k_j=\frac{n_j}{L_j},
\qquad
n_j=0,\ldots,L_j-1 .
\end{align}
Therefore the bond phases entering the Fourier-transformed Hamiltonian are
\begin{align}
\vb k\cdot\bm\delta_x
=2\pi k_1
=\frac{2\pi n_1}{L_1} \equiv q_1,
\quad
\vb k\cdot\bm\delta_y
=2\pi k_2
=\frac{2\pi n_2}{L_2} \equiv q_2,
\quad
\vb k\cdot\bm\delta_z
=0 .
\label{eq:dimensionless-honeycomb-phases}
\end{align}

\subsection{The self-consistent equations for the bond fields}

Equations~\eqref{eq:scf_Kitaev_a} and \eqref{eq:scf_Kitaev_0} define the mean-field amplitudes in terms of the bond fields \(A_a\) and \(B_a\).
The self-consistency loop is completed by computing these bond fields from the correlation functions and substituting them back into Eqs.~\eqref{eq:scf_Kitaev_a} and \eqref{eq:scf_Kitaev_0}.
This procedure corresponds to a uniform, flux-free mean-field ansatz for the Kitaev spin liquid.
At finite temperature, a mean field such as \(B_a\) need not have unit magnitude, so the saddle should not be confused with an exact eigenstate of every \(\mathbb Z_2\) link operator.
Within this approximation, thermal effects enter through the fermionic occupation numbers and the self-consistent bond fields.

In terms of the Abrikosov fermions, the bond fields are
\begin{align}
A_a
\equiv
\Big\langle
i \chi^0_{\vb R,0}
\chi^0_{\vb R+\bm{\delta}_a,1}
\Big\rangle
&=
-i
\Big\langle
f_{\vb R,0,\uparrow}
f_{\vb R+\bm{\delta}_a,1,\uparrow}
\Big\rangle
+i
\Big\langle
f_{\vb R,0,\uparrow}^{\dagger}
f_{\vb R+\bm{\delta}_a,1,\uparrow}
\Big\rangle
+\mathrm{h.c.}
\nonumber\\
&=
2\Im
\Big\langle
f_{\vb R,0,\uparrow}
f_{\vb R+\bm{\delta}_a,1,\uparrow}
\Big\rangle
-
2\Im
\Big\langle
f_{\vb R,0,\uparrow}^{\dagger}
f_{\vb R+\bm{\delta}_a,1,\uparrow}
\Big\rangle ,
\label{eq:scf_chi0}
\\
B_x
\equiv
\Big\langle
i \chi^x_{\vb R,0}
\chi^x_{\vb R+\bm{\delta}_x,1}
\Big\rangle
&=
i
\Big\langle
f_{\vb R,0,\downarrow}
f_{\vb R+\vb a_1,1,\downarrow}
\Big\rangle
+i
\Big\langle
f_{\vb R,0,\downarrow}^{\dagger}
f_{\vb R+\vb a_1,1,\downarrow}
\Big\rangle
+\mathrm{h.c.}
\nonumber\\
&=
-2\Im
\Big\langle
f_{\vb R,0,\downarrow}
f_{\vb R+\vb a_1,1,\downarrow}
\Big\rangle
-
2\Im
\Big\langle
f_{\vb R,0,\downarrow}^{\dagger}
f_{\vb R+\vb a_1,1,\downarrow}
\Big\rangle ,
\label{eq:scf_chix}
\\
B_y
\equiv
\Big\langle
i \chi^y_{\vb R,0}
\chi^y_{\vb R+\bm{\delta}_y,1}
\Big\rangle
&=
-i
\Big\langle
f_{\vb R,0,\downarrow}
f_{\vb R+\vb a_2,1,\downarrow}
\Big\rangle
+i
\Big\langle
f_{\vb R,0,\downarrow}^{\dagger}
f_{\vb R+\vb a_2,1,\downarrow}
\Big\rangle
+\mathrm{h.c.}
\nonumber\\
&=
2\Im
\Big\langle
f_{\vb R,0,\downarrow}
f_{\vb R+\vb a_2,1,\downarrow}
\Big\rangle
-
2\Im
\Big\langle
f_{\vb R,0,\downarrow}^{\dagger}
f_{\vb R+\vb a_2,1,\downarrow}
\Big\rangle ,
\label{eq:scf_chiy}
\\
B_z
\equiv
\Big\langle
i \chi^z_{\vb R,0}
\chi^z_{\vb R+\bm{\delta}_z,1}
\Big\rangle
&=
i
\Big\langle
f_{\vb R,0,\uparrow}
f_{\vb R,1,\uparrow}
\Big\rangle
+i
\Big\langle
f_{\vb R,0,\uparrow}^{\dagger}
f_{\vb R,1,\uparrow}
\Big\rangle
+\mathrm{h.c.}
\nonumber\\
&=
-2\Im
\Big\langle
f_{\vb R,0,\uparrow}
f_{\vb R,1,\uparrow}
\Big\rangle
-
2\Im
\Big\langle
f_{\vb R,0,\uparrow}^{\dagger}
f_{\vb R,1,\uparrow}
\Big\rangle .
\label{eq:scf_chiz}
\end{align}

For numerical evaluation of the correlation functions, we diagonalize the Bogoliubov-de Gennes (BdG) matrix,
\begin{align}
\mathcal H^f_{\vb k,\sigma}
=
\mathcal U^f_{\vb k,\sigma}
\operatorname{diag}\left[
\varepsilon_{\vb k,\sigma,m}
\right]
\mathcal U^{f\,\dagger}_{\vb k,\sigma},
\label{eq:signed_BdG_diagonalization}
\end{align}
where the eigenvalues \(\varepsilon_{\vb k,\sigma,m}\) include both positive and negative BdG partners.

With the Fermi-Dirac function
\begin{align}
f_\beta(\varepsilon)
=
\frac{1}{e^{\beta\varepsilon}+1},
\end{align}
the Nambu covariance matrix is
\begin{align}
C^f_{\vb k,\sigma}
\equiv
\left\langle
\psi^f_{\vb k,\sigma}
\psi_{\vb k,\sigma}^{f\dagger}
\right\rangle
=
\mathcal U^f_{\vb k,\sigma}
\operatorname{diag}
\left[
1-f_\beta\left(\varepsilon_{\vb k,\sigma,m}\right)
\right]
\left(\mathcal U^f_{\vb k,\sigma}\right)^{\dagger}.
\label{eq:C_signed_spectrum}
\end{align}
To see why the occupation factor is \(1-f_\beta\), define the signed Nambu eigenmodes
\begin{align}
\Gamma^f_{\vb k,\sigma}
=
\left(\mathcal U^f_{\vb k,\sigma}\right)^\dagger
\psi^f_{\vb k,\sigma}.
\end{align}
For these modes,
\begin{align}
\left\langle
\Gamma^f_{\vb k,\sigma,m}
\Gamma^{f\dagger}_{\vb k,\sigma,n}
\right\rangle
=
\delta_{mn}
\left[1-f_\beta\!\left(\varepsilon_{\vb k,\sigma,m}\right)\right],
\end{align}
and transforming back gives Eq.~\eqref{eq:C_signed_spectrum}.
At zero temperature, \(1-f_\beta(\varepsilon)\) is replaced by \(\Theta(\varepsilon)\), with the usual half-filling convention for exactly zero modes if they occur.

We also define the density matrix
\begin{align}
\left[G^f_{\vb k,\sigma}\right]_{ll'}
\equiv
\left\langle
\left(\psi_{\vb k,\sigma}^{f\dagger}\right)_l
\left(\psi^f_{\vb k,\sigma}\right)_{l'}
\right\rangle,
\qquad
l,l'=0,1,2,3 .
\end{align}
Using the canonical anticommutation relation \(\left\{ \left(\psi^f_{\vb k,\sigma}\right)_l, \left(\psi_{\vb k,\sigma}^{f\dagger}\right)_{l'} \right\} = \delta_{ll'},\) one obtains
\begin{align}
\left[G^f_{\vb k,\sigma}\right]_{ll'}
&=
\delta_{ll'}-
\left\langle
\left(\psi^f_{\vb k,\sigma}\right)_{l'}
\left(\psi^{f\dagger}_{\vb k,\sigma}\right)_l
\right\rangle
=
\delta_{ll'}-
\left[C^f_{\vb k,\sigma}\right]_{l'l},
\end{align}
or, in matrix notation,
\begin{align}
G^f_{\vb k,\sigma}
=
I_4-\left(C^f_{\vb k,\sigma}\right)^{\intercal}.
\label{eq:G_from_C}
\end{align}
The transpose, not the Hermitian conjugate, is required in Eq.~\eqref{eq:G_from_C} because the indices of \(C^f_{\vb k,\sigma}\) are reversed by the anticommutator.

With the canonical ordering in Eq.~\eqref{eq:psi}, the normal and anomalous correlators are (no summation over the repeated spin index here)
\begin{align}
\Big\langle
f_{\vb R,0,\sigma}^{\dagger}
f_{\vb R+\bm{\delta}_a,1,\sigma}
\Big\rangle
&=
\frac{1}{N_0}
\sum_{\vb k}
e^{-i\vb k\cdot\bm{\delta}_a}
\left[
G^f_{\vb k,\sigma}
\right]_{0,1},
\\
\Big\langle
f_{\vb R,0,\sigma}
f_{\vb R+\bm{\delta}_a,1,\sigma}
\Big\rangle
&=
\frac{1}{N_0}
\sum_{\vb k}
e^{-i\vb k\cdot\bm{\delta}_a}
\left[
G^f_{\vb k,\sigma}
\right]_{2,1}.
\end{align}

\subsection{Mean-field free energy for the Kitaev model}

Given the eigenvalues in Eq.~\eqref{eq:signed_BdG_diagonalization}, the mean-field free energy of the Kitaev model is constructed as follows.
The trace formulas below are exact at every temperature for the stated quadratic mean-field Hamiltonian.
Their relation to the physical spin model requires a separate qualification.
The normalization inherited from the Majorana construction reproduces the ground-state energy in the chosen flux-free sector at \(T=0\), whereas the finite-temperature quadratic trace is a mean-field trace in the enlarged fermionic Hilbert space.
It does not by itself perform the exact local gauge projection or sum over thermally excited flux sectors.

The full Brillouin zone is retained.
For every \(\vb k\), the two physical spin labels and two sublattices give four complex spinon orbitals, while Nambu notation gives eight signed eigenvalues.
The factor \(1/2\) in Eq.~\eqref{eq:Hf} removes this Nambu duplication.

After choosing a nonredundant set of positive-energy quasiparticle modes, a canonical Bogoliubov transformation brings the quadratic part in Eq.~\eqref{eq:Hf} to
\begin{align}
\frac12
\sum_{\vb k,\sigma}
\psi_{\vb k,\sigma}^{f\dagger}
\mathcal H^f_{\vb k, \sigma}
\psi^f_{\vb k,\sigma}
=
\sum_{\substack{\vb k,\sigma,m\\ \varepsilon_{\vb k,\sigma,m}>0}}
\varepsilon_{\vb k, \sigma, m}
\left(
\gamma_{\vb k,\sigma,m}^\dagger\gamma_{\vb k,\sigma, m}
-
\frac12
\right).
\end{align}
The \(-\varepsilon_{\vb k, \sigma, m}/2\) term is the zero-point contribution.
It is part of the fermionic quadratic Hamiltonian and should not be dropped.
Every independent zero mode has vanishing Hamiltonian but still has two occupation states; its entropy must therefore be retained in the partition function.
Then the Abrikosov representation of the Kitaev mean-field Hamiltonian is
\begin{align}
\hat H_f
=
N_0 \varepsilon_f
+
\sum_{\substack{\vb k,\sigma,m\\ \varepsilon_{\vb k,\sigma,m}>0}}
\varepsilon_{\vb k, \sigma, m}
\left(
\gamma_{\vb k, \sigma, m}^\dagger\gamma_{\vb k, \sigma, m}
-
\frac12
\right).
\end{align}

For one positive BdG mode, define
\begin{align}
\hat H
=
\varepsilon
\left(
\gamma^\dagger\gamma-\frac12
\right).
\end{align}
The occupation number of this quasiparticle is \(n_\gamma=0,1\), hence
\begin{align}
Z
=
\sum_{n_\gamma=0,1}
\exp\left[-\beta\varepsilon \left(n_\gamma-\frac12\right)\right]
=
\exp\left(\frac{\beta\varepsilon}{2}\right)
+
\exp\left(-\frac{\beta\varepsilon}{2}\right)
=
2\cosh\left(\frac{\beta\varepsilon}{2}\right).
\label{eq:single_mode_partition_function}
\end{align}
Equation~\eqref{eq:single_mode_partition_function} also gives \(Z=2\) at \(\varepsilon=0\).
Therefore the contribution of this one mode to the free energy is
\begin{align}
F
=
-\frac1\beta
\ln\left[
2\cosh\left(\frac{\beta\varepsilon}{2}\right)
\right]
=
-\frac{\varepsilon}{2}
-
\frac1\beta
\ln\left(1+e^{-\beta\varepsilon}\right).
\label{eq:single_mode_free_energy}
\end{align}
The first term is the zero-point energy, and the second is the thermal occupation contribution.

Multiplying the single-mode partition functions and including the constant factor gives
\begin{align}
Z_f
=
\exp\left(-\beta N_0 \varepsilon_f \right)
2^{N_{\mathrm{zm}}}
\prod_{\substack{\vb k,\sigma,m\\ \varepsilon_{\vb k,\sigma,m}>0}}
2\cosh\left(\frac{\beta\varepsilon_{\vb k, \sigma, m}}{2}\right).
\label{eq:full_partition_function_Kitaev}
\end{align}
Consequently,
\begin{equation}
F_f
=
N_0 \varepsilon_f
-
\frac{N_{\mathrm{zm}}}{\beta}\ln 2
-
\frac1\beta
\sum_{\substack{\vb k,\sigma,m\\ \varepsilon_{\vb k,\sigma,m}>0}}
\ln\left[
2\cosh\left(\frac{\beta\varepsilon_{\vb k, \sigma, m}}{2}\right)
\right].
\label{eq:F_positive_spectrum}
\end{equation}

Let \(N_{\mathrm{zm}}\) be the number of independent complex zero-energy quasiparticle modes.
Equivalently,
\begin{align}
N_{\mathrm{zm}}
=
\frac12
\sum_{\vb k,\sigma}
\dim\ker\mathcal H^f_{\vb k,\sigma}.
\label{eq:number-zero-modes}
\end{align}
If no exact zero mode occurs, \(N_{\mathrm{zm}}=0\), and the last two equations reduce to the usual strict-positive-spectrum formulas.
The explicit zero-mode term is needed, for example, on a finite momentum mesh containing a Dirac point.

If the full signed spectrum is used instead, each nonzero positive energy appears together with its negative partner and \(\cosh\) is even.
The two signed zero eigenvectors associated with one independent complex zero mode contribute \(\tfrac12(2\ln2)=\ln2\).
Thus Eq.~\eqref{eq:F_positive_spectrum} is equivalently
\begin{equation}
F_f
=
N_0 \varepsilon_f
-
\frac{1}{2\beta}
\sum_{\vb k,\sigma,m}
\ln\left[
2\cosh\left(\frac{\beta\varepsilon_{\vb k, \sigma,m}}{2}\right)
\right].
\label{eq:F_signed_spectrum}
\end{equation}
Equation~\eqref{eq:F_signed_spectrum} is the signed-spectrum form of the free energy.
The factor \(1/2\) is present because the sum includes both BdG partners.
If only the positive quasiparticle energies are summed, the prefactor is \(1/\beta\), as in Eq.~\eqref{eq:F_positive_spectrum}, and the independent zero modes must be included separately.

At zero temperature,
\begin{align}
\lim_{\beta\to\infty}
\frac1\beta
\ln\left[
2\cosh\left(\frac{\beta\varepsilon}{2}\right)
\right]
=
\frac{|\varepsilon|}{2} .
\end{align}
Therefore
\begin{equation}
F_f(T=0)
=
N_0\varepsilon_f
-
\frac14
\sum_{\vb k,\sigma, m}
|\varepsilon_{\vb k, \sigma,m}| .
\label{eq:F_zero_temperature}
\end{equation}
For numerical calculations, the expression should be evaluated with the stable form
\begin{align}
\ln\left(2\cosh x\right)
=
|x|+
\ln\left(1+e^{-2|x|}\right),
\qquad
x=\frac{\beta\varepsilon}{2},
\label{eq:stable_logcosh_grand_section}
\end{align}
which avoids overflow when \(\beta|\varepsilon|\) is large.

\subsection{Bogoliubov-de Gennes Hamiltonian for the conduction electrons}
Although the conduction electrons are not paired, we use BdG form here because the Kitaev spin liquid is the paired \(f\) fermion state.
This form is necessary for the Kondo-hybridized phase, where the \(c\) and \(f\) fermions are strongly coupled.
With the canonical ordering,
\begin{align}
\psi^c_{\vb k,\sigma} = 
\left(
c_{\vb k,0,\sigma}, c_{\vb k,1,\sigma},
c_{-\vb k,0,\sigma}^\dagger, c_{-\vb k, 1, \sigma}^\dagger
\right)^\intercal .
\end{align}
It is useful first to write the ordinary two-sublattice Bloch Hamiltonian as
\begin{align}
h^c_{\vb k}
=
\begin{pmatrix}
-\mu&t_{\vb k}\\
t_{\vb k}^*&-\mu
\end{pmatrix},
\quad
t_{\vb k}
=
-\sum_{a=x,y,z}t_a
e^{-i\vb k\cdot\bm\delta_a}.
\label{eq:conduction-normal-bloch}
\end{align}
Because the hopping amplitudes are real, \(t_{-\vb k}=t_{\vb k}^*\).
The BdG matrix is therefore
\begin{align}
\mathcal K^c_{\vb k}
&=
\begin{pmatrix}
h^c_{\vb k}&0\\
0&-h^{c\,\intercal}_{-\vb k}
\end{pmatrix}
=
\begin{pmatrix}
-\mu & t_{\vb k} & 0 & 0 \\
t_{\vb k}^* & -\mu & 0 & 0 \\
0 & 0 & \mu & -t_{\vb k} \\
0 & 0 & -t_{\vb k}^* & \mu
\end{pmatrix}.
\label{eq:conduction-BdG-block}
\end{align}
The constant generated by Nambu normal ordering can be seen directly:
\begin{align}
\frac12\sum_{\vb k,\sigma}
\psi_{\vb k,\sigma}^{c\dagger}
\mathcal K^c_{\vb k}
\psi^c_{\vb k,\sigma}
&=
\sum_{\vb k,\sigma}
\begin{pmatrix}
c_{\vb k,0,\sigma}^\dagger&
c_{\vb k,1,\sigma}^\dagger
\end{pmatrix}
h^c_{\vb k}
\begin{pmatrix}
c_{\vb k,0,\sigma}\\
c_{\vb k,1,\sigma}
\end{pmatrix}
+2\mu N_0.
\label{eq:conduction-Nambu-constant}
\end{align}
Consequently, the conduction-electron grand-canonical Hamiltonian is
\begin{align}
\hat K_c
=
\frac12 \sum_{\vb k,\sigma}
\psi_{\vb k,\sigma}^{c\dagger}
\mathcal K^c_{\vb k}
\psi_{\vb k,\sigma}^c
-2\mu N_0.
\end{align}

Let \(E^c_{\vb k,m}\), with \(m=0,\ldots,3\), denote all four signed eigenvalues of \(\mathcal K^c_{\vb k}\).
Since \(\hat K_c\), rather than \(\hat H_t\), is being traced, the thermodynamic quantity is the grand potential,
\begin{align}
\Omega_c
=
-2\mu N_0
-
\frac{1}{\beta}
\sum_{\vb k,m}
\ln\left[
2\cosh\left(\frac{\beta E^c_{\vb k,m}}{2}\right)
\right].
\label{eq:conduction-grand-potential-signed}
\end{align}
The spin degeneracy cancels the factor \(1/2\) that appears in Eq.~\eqref{eq:F_signed_spectrum}.
Equivalently, in terms of the two ordinary band energies
\begin{align}
\xi_{\vb k,\eta}
=
\eta\left|t_{\vb k}\right|-\mu,
\qquad
\eta=\pm,
\end{align}
one obtains the familiar normal-state expression
\begin{align}
\Omega_c
=
-\frac{2}{\beta}
\sum_{\vb k,\eta=\pm}
\ln\left(1+e^{-\beta\xi_{\vb k,\eta}}\right).
\label{eq:conduction-grand-potential-normal}
\end{align}
Equations~\eqref{eq:conduction-grand-potential-signed} and \eqref{eq:conduction-grand-potential-normal} are identical; the constant in Eq.~\eqref{eq:conduction-Nambu-constant} is what makes the two forms agree.

\section{Kondo-hybridized superconductor}
\label{sec:SC}

\subsection{Abrikosov fermion representation of the Kondo coupling}

We now rewrite the onsite Kondo exchange in singlet hopping and pairing channels.
In this subsection the site index is suppressed.
The \(c\) and \(f\) operators are taken to belong to one fermionic algebra, so all mixed anticommutators vanish, for example \(\{c_\alpha,f_\beta\}=0\) and \(\{c_\alpha,f_\beta^\dagger\}=0\).
Define
\begin{align}
\hat V
=
c^\dagger_\alpha f_\alpha,
\quad
\hat\Delta
=
c_\alpha \epsilon_{\alpha\beta} f_\beta
=
c_\uparrow f_\downarrow-c_\downarrow f_\uparrow,
\quad
\epsilon=i\sigma^y,
\end{align}
Both bilinears are spin singlets.
Under a common spin rotation \(c\mapsto Uc\), \(f\mapsto Uf\), one uses \(U^\dagger U=I_2\) for \(\hat V\) and \(U^\intercal\epsilon U=\epsilon\) for \(\hat\Delta\).
The onsite Kondo-pairing field \(\Delta\) introduced here is distinct from the superscripted spinon bond coefficients \(\Delta^a_{\vb k}\) in Eqs.~\eqref{eq:scf_Kitaev_a} and \eqref{eq:scf_Kitaev_0}.
We also define
\begin{align}
\hat n_c=c^\dagger_\alpha c_\alpha,
\qquad
\hat n_f=f^\dagger_\alpha f_\alpha,
\qquad
\hat Q=c^\dagger_\alpha c_\beta f^\dagger_\beta f_\alpha .
\end{align}
Using \(\sigma^a_{\alpha\beta}\sigma^a_{\gamma\delta} = 2\delta_{\alpha\delta}\delta_{\beta\gamma} - \delta_{\alpha\beta}\delta_{\gamma\delta},\) the exchange operator is
\begin{align}
\hat{\bm s}\cdot \hat{\bm S}
&=
\frac14
c^\dagger_\alpha \sigma^a_{\alpha\beta}c_\beta
f^\dagger_\gamma \sigma^a_{\gamma\delta}f_\delta
=
\frac12 \hat Q-
\frac14 \hat n_c\hat n_f .
\label{eq:kondo-exchange-Q}
\end{align}
The singlet hopping bilinear gives
\begin{align}
\hat V^\dagger \hat V
&=
f^\dagger_\alpha c_\alpha c^\dagger_\beta f_\beta
=
f^\dagger_\alpha
\left(\delta_{\alpha\beta}-c^\dagger_\beta c_\alpha\right)
f_\beta
=
\hat n_f-
\hat Q .
\label{eq:kondo-hopping-identity}
\end{align}
In the last step, moving \(f_\alpha^\dagger\) through \(c_\beta^\dagger c_\alpha\) produces two minus signs, and relabeling \(\alpha\leftrightarrow\beta\) gives
\begin{align}
f_\alpha^\dagger c_\beta^\dagger c_\alpha f_\beta
=
c_\beta^\dagger c_\alpha f_\alpha^\dagger f_\beta
=
\hat Q.
\end{align}
For the pairing bilinear, the identity \(\epsilon_{\alpha\beta}\epsilon_{\gamma\delta} = \delta_{\alpha\gamma}\delta_{\beta\delta} - \delta_{\alpha\delta}\delta_{\beta\gamma}\) gives
\begin{align}
\hat\Delta^\dagger \hat\Delta
&=
f^\dagger_\beta \epsilon_{\alpha\beta}
 c^\dagger_\alpha c_\gamma
\epsilon_{\gamma\delta} f_\delta
=
c^\dagger_\alpha c_\gamma
f^\dagger_\beta f_\delta
\epsilon_{\alpha\beta}\epsilon_{\gamma\delta}
=
\hat n_c\hat n_f-
\hat Q .
\label{eq:kondo-pairing-identity}
\end{align}
Here,
\begin{align}
f_\beta^\dagger c_\alpha^\dagger c_\gamma f_\delta
=
c_\alpha^\dagger c_\gamma f_\beta^\dagger f_\delta,
\end{align}
because the two exchanges of an \(f\) operator with a \(c\) operator cancel.
Combining Eqs.~\eqref{eq:kondo-hopping-identity} and \eqref{eq:kondo-pairing-identity},
\begin{align}
\hat V^\dagger\hat V
+
\hat\Delta^\dagger\hat\Delta
=
\hat n_f+
\hat n_c\hat n_f
-
2\hat Q .
\end{align}
Thus
\begin{align}
\hat Q
=
\frac12
\left(
\hat n_f+\hat n_c\hat n_f
-\hat V^\dagger\hat V
-\hat\Delta^\dagger\hat\Delta
\right).
\end{align}
Substituting this result into Eq.~\eqref{eq:kondo-exchange-Q} cancels the \(\hat n_c\hat n_f\) terms and gives
\begin{align}
\hat{\bm s}\cdot \hat{\bm S}
=
-
\frac14
\left(
\hat V^\dagger\hat V
+
\hat\Delta^\dagger\hat\Delta
\right)
+
\frac14\hat n_f .
\label{eq:kondo-singlet-decomposition}
\end{align}
In the physical Abrikosov-fermion Hilbert space, \(\hat n_f=1\), so the last term is field independent and is omitted henceforth.
The Kondo interaction used below is therefore
\begin{align}
\hat H_{\mathrm{Kondo}}
=
-
\frac{J_K}{4}\sum_j
\left(
\hat V_j^\dagger\hat V_j
+
\hat\Delta_j^\dagger\hat\Delta_j
\right).
\label{eq:kondo-singlet-H}
\end{align}

For antiferromagnetic \(J_K>0\), both singlet channels are attractive.
For either \(\hat O=\hat V\) or \(\hat O=\hat\Delta\), write \(\hat O=O+\delta\hat O\).
Then
\begin{align}
\hat O^\dagger\hat O
=
O^*\hat O+O\hat O^\dagger-|O|^2
+\delta\hat O^\dagger\delta\hat O.
\end{align}
The saddle-point approximation drops the last, quadratic fluctuation term.
Equivalently, this is obtained after a Hubbard--Stratonovich transformation.
One obtains
\begin{align}
\hat H_{\mathrm{Kondo}}^{\rm MF} &\equiv \hat H_{cf}
=
-
\frac{J_K}{4}\sum_j
\left[
V_j\hat V_j^\dagger
+
V_j^*\hat V_j
+
\Delta_j\hat\Delta_j^\dagger
+
\Delta_j^*\hat\Delta_j
-
|V_j|^2
-
|\Delta_j|^2
\right]
\nonumber\\
&=
-
\frac{J_K}{4}
\sum_j
\Bigl[
V_j^* c^\dagger_{j\alpha} f_{j\alpha}
+
V_j f^\dagger_{j\alpha} c_{j\alpha}
+
\Delta_j^*
\left(c_{j\uparrow} f_{j\downarrow}-c_{j\downarrow} f_{j\uparrow}\right)
+
\Delta_j
\left(f^\dagger_{j\downarrow} c^\dagger_{j\uparrow}
-
f^\dagger_{j\uparrow} c^\dagger_{j\downarrow}\right)
\Bigr]
\nonumber \\
&\quad+
\frac{J_K}{4}\sum_j
\left(|V_j|^2+|\Delta_j|^2\right).
\label{eq:kondo-MF-expanded}
\end{align}
The self-consistent conditions, obtained by varying with respect to \(V_j^*\) and \(\Delta_j^*\), are
\begin{align}
V_j&=\ev{\hat V_j} = \ev{c_{j\sigma}^\dagger f_{j\sigma}},
\\
\Delta_j&=\ev{ \hat\Delta_j} = \ev{c_{j\uparrow} f_{j\downarrow}-c_{j\downarrow} f_{j\uparrow}}.
\end{align}

We assume translationally invariant condensates, while allowing the two sublattices to differ.
Thus,
\begin{align}
V_j = V_{\vb R, s} = V_s,
\qquad
\Delta_j = \Delta_{\vb R,s} =\Delta_s,
\qquad
s=0,1.
\end{align}

The one-spinon constraint in Eq.~\eqref{eq:one-spinon-constraint} must also be enforced at the saddle.
In the isolated, particle--hole-symmetric FL* state, the average constraint is satisfied by a vanishing static multiplier.
Once the \(c\) and \(f\) sectors hybridize, that property is no longer automatic.
We therefore introduce real, sublattice-dependent Lagrange multipliers \(\lambda_s\) through
\begin{align}
\hat H_\lambda
&=
\sum_{\vb R}
\sum_{s=0,1}
\lambda_s
\left(
\sum_{\sigma}
f_{\vb R,s,\sigma}^\dagger f_{\vb R,s,\sigma}
-1
\right).
\label{eq:spinon-constraint-lagrange}
\end{align}
This is a saddle-point enforcement of the local constraint on average, rather than an exact site-by-site projection of the thermal trace.
In the four-component spinon Nambu basis, define
\begin{align}
\mathcal L
&=
\operatorname{diag}
\left(
\lambda_0,\lambda_1,-\lambda_0,-\lambda_1
\right).
\label{eq:spinon-constraint-BdG-block}
\end{align}
Summing \(\tfrac12\psi^{f\dagger}\mathcal L\psi^f\) over both physical spin blocks gives Eq.~\eqref{eq:spinon-constraint-lagrange}, including its \(-N_0\sum_s\lambda_s\) term.
No additional \(\lambda_s\)-dependent constant should therefore be added below.

\subsection{Mean-field Hamiltonian}
\label{ssec:Hmf}

Together with the Kondo interaction and the constraint term, we can now write the mean-field grand-canonical Hamiltonian for the total system in the Nambu basis:
\begin{align}
\hat K_{\text{MF}} &= \frac12 \sum_{\vb k}
\left(
\psi_{\vb k,\uparrow}^{c \,\dagger},
\psi_{\vb k,\uparrow}^{f\,\dagger},
\psi_{\vb k,\downarrow}^{c\,\dagger},
\psi_{\vb k,\downarrow}^{f\,\dagger}
\right)
\begin{pmatrix}
\mathcal{K}_{\vb k}^c & \mathcal V & 0 & \mathcal P_K
\\
\mathcal V^\dagger & \mathcal{H}_{\vb k,\uparrow}^f+\mathcal L & \mathcal P_K & 0
\\
0 & \mathcal P_K^\dagger & \mathcal{K}_{\vb k}^c & \mathcal V
\\
\mathcal P_K^\dagger& 0 & \mathcal V^\dagger &\mathcal{H}_{\vb k,\downarrow}^f+\mathcal L
\end{pmatrix}
\begin{pmatrix}
\psi_{\vb k,\uparrow}^c \\
\psi_{\vb k,\uparrow}^f \\
\psi_{\vb k,\downarrow}^c \\
\psi_{\vb k,\downarrow}^f
\end{pmatrix}
+ N_0 \varepsilon_{\text{MF}}
\nonumber\\
&=\frac12 \sum_{\vb k} \Psi_{\vb k}^\dagger
\mathcal H_{\text{MF}}(\vb k) \Psi_{\vb k}
+ N_0 \varepsilon_{\text{MF}}.
\label{eq:full-mean-field-Hamiltonian}
\end{align}
where
\begin{align}
\mathcal V = -\frac{J_K}4
\begin{pmatrix}
V_0^* & 0 & 0 & 0 \\
0 & V_1^* & 0 & 0 \\
0 & 0 & -V_0 & 0 \\
0 & 0 & 0 & -V_1
\end{pmatrix},
\quad
\mathcal P_K = -\frac{J_K}4
\begin{pmatrix}
0 & 0 & -\Delta_0 & 0 \\
0 & 0 & 0 & -\Delta_1 \\
\Delta_0^* & 0 & 0 & 0 \\
0 & \Delta_1^* & 0 & 0
\end{pmatrix},
\end{align}
and
\begin{align}
\varepsilon_{\text{MF}}
=
-2\mu
+\varepsilon_f
+\frac{J_K}4
\sum_{s=0,1}
\left( |V_s|^2 + |\Delta_s|^2 \right).
\label{eq:mean-field-constant}
\end{align}
The \(-2\mu\) term is the conduction-electron Nambu constant per unit cell, and the last term is the Hubbard--Stratonovich mean-field cost.
The Nambu spinor is defined as
\begin{align}
\Psi_{\vb k}^\intercal \equiv
\left(
\psi_{\vb k,\uparrow}^{c \,\intercal},
\psi_{\vb k,\uparrow}^{f\,\intercal},
\psi_{\vb k,\downarrow}^{c\,\intercal},
\psi_{\vb k,\downarrow}^{f\,\intercal}
\right).
\end{align}

For completeness, the onsite Kondo terms in momentum space are
\begin{align}
\hat H_{cf}^{\mathrm{bilinear}}
&=
-\frac{J_K}{4}
\sum_{\vb k,s}
\Bigl[
V_s^*
\left(
c_{\vb k,s,\uparrow}^\dagger f_{\vb k,s,\uparrow}
+c_{\vb k,s,\downarrow}^\dagger f_{\vb k,s,\downarrow}
\right)
+\Delta_s^*
\left(
c_{\vb k,s,\uparrow}f_{-\vb k,s,\downarrow}
-c_{\vb k,s,\downarrow}f_{-\vb k,s,\uparrow}
\right)
+\mathrm{h.c.}
\Bigr].
\label{eq:kondo-bilinear-momentum}
\end{align}
The minus signs in the hole sector of \(\mathcal V\), and the relative signs in \(\mathcal P_K\), follow from reordering the fermions in Eq.~\eqref{eq:kondo-bilinear-momentum}.
This provides a direct check that the block matrix in Eq.~\eqref{eq:full-mean-field-Hamiltonian} is Hermitian and has the required antisymmetric pairing structure.

\subsection{Correlation functions and the self-consistent equations}

When either \(V_s\) or \(\Delta_s\) is finite, the full mean-field matrix is needed to calculate every correlation function.
We diagonalize
\begin{align}
\mathcal H_{\text{MF}}(\vb k)
&=
\mathcal U_{\vb k}
\operatorname{diag}
\left[
E_{\vb k,0},\ldots,E_{\vb k,15}
\right]
\mathcal U^\dagger_{\vb k}.
\label{eq:full-signed-diagonalization}
\end{align}
The sixteen eigenvalues at each momentum are all retained.
The collection over \((\vb k,m)\) contains the signed BdG partners, with a partner generally allowed to occur at \(-\vb k\).

The covariance matrix is
\begin{align}
C_{\vb k}
&\equiv
\left\langle\Psi_{\vb k}\Psi_{\vb k}^\dagger\right\rangle
=
\mathcal U_{\vb k}
\operatorname{diag}
\left[
1-f_{\beta}\left(E_{\vb k,m}\right)
\right]
\mathcal U_{\vb k}^\dagger,
\end{align}
where the diagonal expression is understood for \(m=0,\ldots,15\).
As in Eq.~\eqref{eq:G_from_C}, the corresponding density matrix is
\begin{align}
\left[G_{\vb k}\right]_{ll'}
&\equiv
\left\langle
\Psi^\dagger_{\vb k,l}\Psi_{\vb k,l'}
\right\rangle
=
\delta_{ll'}-\left[C_{\vb k}\right]_{l'l},
\qquad
l,l'=0,\ldots,15,
\nonumber\\
G_{\vb k}
&=
I_{16}-C_{\vb k}^{\intercal}.
\label{eq:full-G-from-C}
\end{align}

For the component indices below, we identify \(\sigma=\uparrow=0\) and \(\sigma=\downarrow=1\).

The correlation functions are then as follows.
There is no summation over \(\sigma\) in the first two lines; every spin sum is displayed explicitly:
\begin{align}
\Big\langle
f_{\vb R,0,\sigma}^{\dagger}
f_{\vb R+\bm{\delta}_a,1,\sigma}
\Big\rangle
&=
\frac{1}{N_0}
\sum_{\vb k}
e^{-i\vb k\cdot\bm{\delta}_a}
\left[
G_{\vb k}
\right]_{8\sigma+4,\,8\sigma+5},
\label{eq:full-spinon-normal-correlator}
\\
\Big\langle
f_{\vb R,0,\sigma}
f_{\vb R+\bm{\delta}_a,1,\sigma}
\Big\rangle
&=
\frac{1}{N_0}
\sum_{\vb k}
e^{-i\vb k\cdot\bm{\delta}_a}
\left[
G_{\vb k}
\right]_{8\sigma+6,\,8\sigma+5},
\label{eq:full-spinon-anomalous-correlator}
\\
V_s
=
\sum_{\sigma}
\Big\langle
c_{\vb R,s,\sigma}^\dagger f_{\vb R,s,\sigma}
\Big\rangle
&=
\frac1{N_0}
\sum_{\vb k,\sigma}
\left[
G_{\vb k}
\right]_{8\sigma+s,\,8\sigma+4+s},
\label{eq:full-V-self-consistency}
\\
\Delta_s
=
\Big\langle
c_{\vb R,s,\uparrow} f_{\vb R,s,\downarrow}
-c_{\vb R,s,\downarrow} f_{\vb R,s,\uparrow}
\Big\rangle
&=
\frac1{N_0}
\sum_{\vb k}
\left(
\left[G_{\vb k}\right]_{s+2,\,s+12}
-
\left[G_{\vb k}\right]_{s+10,\,s+4}
\right).
\label{eq:full-kondo-self-consistency}
\end{align}

Variation with respect to the two constraint fields gives the additional self-consistency equations
\begin{align}
1
&=
\frac1{N_0}
\sum_{\vb k,\sigma}
\left[G_{\vb k}\right]_{
8\sigma+4+s,\,8\sigma+4+s},
\qquad
s=0,1.
\label{eq:spinon-number-constraint-full}
\end{align}
Equations~\eqref{eq:scf_chi0}--\eqref{eq:scf_chiz}, evaluated with the full correlators in Eqs.~\eqref{eq:full-spinon-normal-correlator} and \eqref{eq:full-spinon-anomalous-correlator}, together with Eqs.~\eqref{eq:full-V-self-consistency}--\eqref{eq:spinon-number-constraint-full}, form the complete grand-canonical saddle-point equations for \(A_a,B_a,V_s,\Delta_s,\lambda_s\).

If \(\mu\) is prescribed, no further equation is required.
At fixed conduction-electron density per honeycomb site, including both physical spin states, define
\begin{align}
n_c
&\equiv
\frac{1}{2N_0}
\sum_{\vb R,s,\sigma}
\left\langle
c_{\vb R,s,\sigma}^\dagger c_{\vb R,s,\sigma}
\right\rangle
=
\frac{1}{2N_0}
\sum_{\vb k,s,\sigma}
\left[G_{\vb k}\right]_{
8\sigma+s,\,8\sigma+s},
\qquad
0\le n_c\le2,
\label{eq:full-conduction-density}
\end{align}
and determine \(\mu\) from this equation.

Because Eq.~\eqref{eq:full-mean-field-Hamiltonian} contains \(-\mu\hat N_c\), its signed-spectrum trace is a grand potential, not a Helmholtz free energy.
The grand potential per unit cell is
\begin{align}
\omega
\equiv
\frac{\Omega}{N_0}
&=
\varepsilon_{\text{MF}}
-\frac1{2\beta N_0}
\sum_{\vb k,m=0}^{15}
\ln\left[
2\cosh\left(\frac{\beta E_{\vb k,m}}2\right)
\right].
\label{eq:full-grand-potential}
\end{align}
The factor \(1/2\) is required because all signed BdG eigenvalues are summed.
Exact zero modes are included automatically.
The stable identity in Eq.~\eqref{eq:stable_logcosh_grand_section} should be used for each term.
Finally, at fixed \(n_c\), the Helmholtz free energy per unit cell is the Legendre transform
\begin{align}
f(n_c)
&=
\omega+2\mu n_c,
\label{eq:fixed-density-legendre-transform}
\end{align}
because \(N_c=2N_0n_c\).

\subsection{Superconducting and time-reversal-odd observables}

We now define the physical observables shown in Fig.~1 of the main
text.  The onsite Kondo hybridization and pairing fields form the
Kondo doublet
\begin{equation}
h_s =
\begin{pmatrix}
V_s\\
-\Delta_s^*
\end{pmatrix},
\qquad s=0,1 .
\label{eq:Kondo_doublet_SM}
\end{equation}
The Kondo amplitude plotted in Fig.~1(a) is
\begin{equation}
\bar h =
\frac{1}{2}\sum_{s=0,1}
\sqrt{|V_s|^2+|\Delta_s|^2}.
\label{eq:hbar_SM}
\end{equation}
For the sublattice-related solution discussed in the main text,
the two local norms are equal, so $\bar h=\|h_0\|=\|h_1\|$.

\emph{Charge-two superconducting order.---}
The superconducting order generated by Kondo condensation is
diagnosed by the gauge-invariant charge-two bond composite
\begin{equation}
\mathcal P^\mu_{jk}
=
h_j^T u^\mu_{jk}\epsilon h_k,
\qquad
\epsilon =
\begin{pmatrix}
0&1\\
-1&0
\end{pmatrix},
\label{eq:P_def_SM}
\end{equation}
where $u^\mu_{jk}$ is the spinon mean-field matrix on the oriented
bond $j=(\mathbf R,0)$,
$k=(\mathbf R+\boldsymbol\delta_a,1)$.
Here $\mu=0$ denotes the spin-singlet channel and
$\mu=x,y,z$ the spin-triplet channels.

For the Kitaev mean-field ansatz used above, these matrices can be
written as
\begin{align}
u^0_{jk}
&=
i\alpha_a^0 I_2,
&
\alpha_a^0
&=
\frac{K_a}{8}(B_a+A_a),
\label{eq:u0_SM}
\\
u^b_{jk}
&=
i\alpha_a^b\tau_b,
&
\alpha_a^b
&=
\frac{K_a}{8}
\left[
B_a+(2\delta_{ab}-1)A_a
\right],
\qquad b=x,y,z ,
\label{eq:ub_SM}
\end{align}
where $\tau_b$ are Pauli matrices acting in the internal
spinon gauge/Nambu space.  The matrices $u^\mu_{jk}$ are
Hamiltonian coefficients determined by the self-consistent bond
fields $A_a$ and $B_a$; they should not be confused with
equal-time spinon correlation matrices.

For $h_j=(V_j,-\Delta_j^*)^T$, Eq.~\eqref{eq:P_def_SM} gives
\begin{align}
\mathcal P^0_{jk}
&=
i\alpha_a^0
\left(
\Delta_j^* V_k-V_j\Delta_k^*
\right),
\label{eq:P0_SM}
\\
\mathcal P^x_{jk}
&=
i\alpha_a^x
\left(
\Delta_j^*\Delta_k^*-V_jV_k
\right),
\label{eq:Px_SM}
\\
\mathcal P^y_{jk}
&=
-\alpha_a^y
\left(
V_jV_k+\Delta_j^*\Delta_k^*
\right),
\label{eq:Py_SM}
\\
\mathcal P^z_{jk}
&=
-i\alpha_a^z
\left(
V_j\Delta_k^*+\Delta_j^*V_k
\right).
\label{eq:Pz_SM}
\end{align}
We denote the value on an $a$-type bond by
$\mathcal P_a^\mu\equiv
\mathcal P^\mu_{\mathbf R0,\mathbf R+\boldsymbol\delta_a,1}$.
The quantity plotted in Fig.~1(b) of the main text is
$\mathcal P_a^y$.
Under the physical electromagnetic transformation
$c_j\rightarrow e^{i\theta}c_j$,
\begin{equation}
\mathcal P^\mu_{jk}
\rightarrow
e^{-2i\theta}\mathcal P^\mu_{jk},
\end{equation}
so a nonzero $\mathcal P^\mu_{jk}$ is a charge-two
superconducting order parameter.

\emph{Physical electronic pair amplitude and time-reversal breaking.---}
Independently of the composite order parameter above, we calculate
the anomalous conduction-electron correlator directly from the
full self-consistent density matrix,
\begin{equation}
\Phi_a^{\sigma\sigma'}
\equiv
\left\langle
c_{\mathbf R,0,\sigma}
c_{\mathbf R+\boldsymbol\delta_a,1,\sigma'}
\right\rangle
=
\frac{1}{N_0}
\sum_{\mathbf k}
e^{-i\mathbf k\cdot\boldsymbol\delta_a}
[G_{\mathbf k}]_{8\sigma+2,\,8\sigma'+1},
\label{eq:Phi_SM}
\end{equation}
where $\sigma=\uparrow,\downarrow$ is identified with
$0,1$, respectively, in the matrix indices of Eq.~(S99).

We decompose the $2\times2$ pair matrix as
\begin{equation}
\Phi_a
=
\left(
\psi_a+\mathbf d_a\cdot\boldsymbol\sigma
\right)i\sigma_y ,
\label{eq:pair_decomp_SM}
\end{equation}
with
\begin{align}
\psi_a
&=
\frac{
\Phi_a^{\uparrow\downarrow}
-
\Phi_a^{\downarrow\uparrow}
}{2},
\label{eq:psi_SM}
\\
d_a^x
&=
\frac{
\Phi_a^{\downarrow\downarrow}
-
\Phi_a^{\uparrow\uparrow}
}{2},
\label{eq:dx_SM}
\\
d_a^y
&=
\frac{
\Phi_a^{\uparrow\uparrow}
+
\Phi_a^{\downarrow\downarrow}
}{2i},
\label{eq:dy_SM}
\\
d_a^z
&=
\frac{
\Phi_a^{\uparrow\downarrow}
+
\Phi_a^{\downarrow\uparrow}
}{2}.
\label{eq:dz_SM}
\end{align}
The nonunitary triplet-pairing vector is
\begin{equation}
\mathbf q_a
=
i\,\mathbf d_a\times\mathbf d_a^* .
\label{eq:q_SM}
\end{equation}
The vector $\mathbf q_a$ is real and invariant under an overall
electromagnetic phase rotation of the pair amplitude.  A nonzero
$\mathbf q_a$ is therefore a sufficient physical diagnostic of
time-reversal-symmetry breaking.  For the self-consistent solution
shown in Fig.~1(c) of the main text,
\begin{equation}
\mathbf q_a=(q_a^x,0,0),
\end{equation}
and $q_a^x$ is the quantity plotted there.

Thus, the three quantities shown in Fig.~1 probe distinct aspects
of the same transition: $\bar h$ measures the onset of Kondo
condensation, $\mathcal P_a^y$ diagnoses physical charge-two
superconducting order, and $q_a^x$ diagnoses time-reversal-symmetry
breaking.  Within the numerical resolution specified below, all
three become nonzero at the same critical temperature $T_c$.